\documentclass[twocolumn]{revtex4}

\usepackage{epsfig}
\usepackage{amsmath}
\usepackage{bm}

\begin{document}

\title{A practical limitation for continuous-variable quantum cryptography using coherent states}
\author{Ryo Namiki}
\email[Electric address: ]{namiki@qo.phys.gakushuin.ac.jp} 
\author{Takuya Hirano }
 
\affiliation{CREST Research Team for Photonic Quantum Information, Department of Physics, Gakushuin University,
Mejiro 1-5-1, Toshima-ku, Tokyo 171-8588, Japan}
\date{\today}
\begin{abstract}
In this letter, first, we investigate the security of a continuous-variable quantum cryptographic scheme with a postselection process against individual beam splitting attack. It is shown that the scheme can be secure in the presence of the transmission loss owing to the postselection.
Second, we provide a loss limit for continuous-variable quantum cryptography using coherent states taking into account excess Gaussian noise on quadrature distribution.
Since the excess noise is reduced by the loss mechanism, a realistic intercept-resend attack which makes a Gaussian mixture of coherent states gives a loss limit in the presence of any excess Gaussian noise.
\end{abstract}

\maketitle


The security of quantum cryptography is degraded by the presence of realistic experimental imperfections. In particular the transmission loss limits the performance of schemes for a long distance transmission \cite{br1}.

Recently several continuous-variable quantum cryptographic schemes have been proposed \cite{squeezed,epr,con,alcon,coherent,postsel,hirano,namiki1}. Those are sorted into either all-continuous type or hybrid type \cite{alcon}, the all-continuous scheme distributes a continuous key and the hybrid scheme distributes a discrete key. A loss limit, in the sense that the mutual information between Alice and Bob $I_{AB}$ cannot be greater than the Shannon information of an Eavesdropper (Eve) $I_E$, is given for an all-continuous scheme \cite{coherent} and it is shown that this limitation can be removed by introducing a postselection process for a hybrid scheme \cite{postsel,hirano,namiki1}. The existence of loss limit is an open question.

 The reliable security measure for discrete quantum cryptographic schemes against individual attacks is the secure key gain $G$ which ensures that 
$I_E$ can be arbitrarily small in the long key limit if $G$ is positive \cite{l3,l2}. The question is how high $G$ can be for a given loss or transmission distance in realistic conditions. The estimations are given for BB84 protocol \cite{l2}, entangled photon protocol \cite{entangle-photons}, and B92 protocol \cite{kiyotama1}. The estimation of $G$ for continuous schemes, if possible, is important as a comparison with discrete schemes. At least, the framework \cite{l1,A57} can be adapted to hybrid schemes.

For these discrete schemes, the experimental imperfections are mostly determined by observed bit error rate and dark count rate of single photon detectors \cite{l2,entangle-photons,kiyotama1}.
In continuous-variable schemes, the experimental imperfections appear as the change of quadrature distributions. 
Experimentally, quadrature measurement is performed slightly above the standard quantum limit and observed quadrature distribution has additional Gaussian noise upon the minimum uncertainty Gaussian wavepacket \cite{hirano}.   
Thus, the security analysis including experimental imperfections seems to become qualitatively different from that of the discrete schemes.

In our previous work \cite{namiki1} we estimated $G$ of a hybrid type scheme applying a postselection \cite{hirano} for a given loss, provided Eve performs quadrature measurement for the lost part of the signal. In this case it is shown that $G$ can be positive if the loss is less than unity by setting a large postselection threshold.

In this letter, firstly we estimate $G$ of this scheme \cite{hirano} for a given loss  against individual beam splitting attack \cite{postsel}, that is, Eve can use a positive operator valued measure (POVM) on the individual split signal independently. It is shown that $G$ can be positive by setting a large threshold and there is no loss limit.
 Secondly we provide two concrete examples of eavesdropping attack which causes excess Gaussian noise. The first one is treated as an extension of individual beam splitting attack and it also shows no loss limit. The second one is an intercept-resend attack. It imposes a practical loss limit on every coherent state scheme.


The protocol we study here is a four state protocol using phase modulation of weak coherent pulse and balanced homodyne detection applying a postselection process \cite{hirano,namiki1}.
Alice randomly chooses one of the four coherent states $|\sqrt{n} e^{i m\pi/2} \rangle $ with the pulse intensity (the mean photon number per pulse) $n>0$, $m=0, 1, 2, 3   $ and sends it to Bob. 
Then Bob randomly measures one of the two quadratures $\hat{x}_k$ with $k =1,2 $ and $[\hat x_1 , \hat x_2]=\frac{i}{2} $. 
After the transmission of a large number of pulses, Alice transmits the parity of $m$ to Bob through a classical channel. For the pulses $m-k =\pm 1$, Bob sets a threshold $x_0 (\ge 0)$ and constructs his bit sequence by the following decision:
\begin{eqnarray}
({\text{bit  value}}) = \left\{ 
	\begin{array}{ll}
           1  & {\text{if }}  x  >  x_0  \\
            0  &{\text{if }}     x   <-x_0,
\end{array} \right.  \label{deceq}
\end{eqnarray} where $x$ is the result of Bob's measurement. This is a postselection process and the advantage is that Bob can obtain arbitrarily small bit error rate by setting a larger threshold in the absence of obvious eavesdropping \cite{namiki1}. For simplicity, hereafter we set $k=1$. This does not change the results of following discussion.

\begin{figure}[htbp]
\epsfig{file=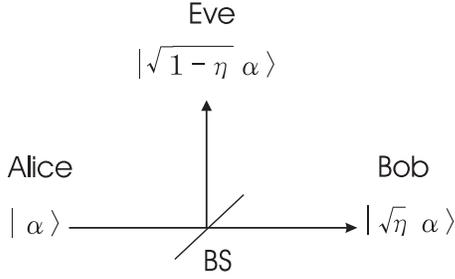,width=6cm}
\caption{ \label{fig1} 
In a lossy channel, the input coherent state $| \alpha \rangle $ is simply amplitude-damped as $| \sqrt{\eta} \alpha \rangle$ 
by the loss $1-\eta$.
In beam splitting attack, Eve replaces the lossy transmission path with her lossless one and uses a beam splitter (BS) with the reflectivity $1-\eta $, which corresponds to the original transmission loss. Then Eve can obtain the lost part of the signal $| \sqrt{1-\eta} \alpha \rangle$ without causing any further disturbance to Bob's signal. }
\end{figure}

The lossy channel is modeled by a beam splitter (see Fig. \ref{fig1}) and the split signal is assumed to be received by Eve.
 Bob receives the signal $|\sqrt{\eta n}  e^{i m\pi/2} \rangle$, where $\eta  \ ( 0 < \eta \le 1)$ is the parameter characterizing the loss $1-\eta$. If $m-k = \pm 1$, the probability that Bob's quadrature measurement results $x$ is given by \cite{namiki1}:
\begin{widetext}
\begin{eqnarray}
 \textrm{Prob}(x) 
& =& \frac{1}{2} \left(  |\langle x_1| \sqrt{\eta n} \rangle|^2+|\langle x_1| -\sqrt{\eta n} \rangle|^2 \right) \Big{|} _{x_1=x} \nonumber\\
     & =&\frac{1}{2} \frac{1}{ \sqrt{2 \pi (\Delta x )^2 }}\left\{ \exp{ \left[ -\frac{ (x-\sqrt{\eta n} ) ^2}{2(\Delta x)^2 } \right] }+ \exp{\left[ -\frac{  (x+ \sqrt{\eta n} ) ^2}{2 (\Delta x )^2} \right] } \right\}, \label{pd}  
\end{eqnarray} 
 \end{widetext}
  where $\langle x_1|$ denotes the eigenbra of $\hat x_1 $ and $(\Delta x )^2 = \frac{1}{4}$ is the quadrature variance of coherent state.
Using this probability, Bob's Shannon information gain per pulse is given by \cite{A57}: 
\begin{eqnarray}
 \frac{1}{2 }\sum_{|x|>x_0}\textrm{Prob}(x) i_{AB} ( x,\eta n ), \label{s-gain}
\end{eqnarray}
where the factor 1/2 is the probability that the basis is correct, i.e., Bob's choice of $k$ satisfies $m-k = \pm 1$, and   

\begin{eqnarray}
i_{AB}(x,n) &=&  1+ \textrm{Prob}(\sqrt{n} |x) \log _2{\textrm{Prob}(\sqrt{n} |x)} \nonumber \\&+&\textrm{Prob}(-\sqrt{n} |x) \log _2{\textrm{Prob}(-\sqrt{n} |x)} 
\end{eqnarray}
is the Shannon information gain when $x$ is triggered, where \begin{eqnarray}
{\textrm{ Prob}}(\sqrt{n} |x_1)&=&\frac{\left| \langle x_1|\sqrt{n}\rangle \right|^2 }{ \left|\langle x_1|\sqrt{n} \rangle \right|^2+\left| \langle  x_1|-\sqrt{n} \rangle \right|^2 }\nonumber\\
&=& \frac{1}{ 1+\exp \left[ -\frac{2 \sqrt{n} x_1}{ (\Delta x )^2} \right] } \label{replace}
\end{eqnarray}
is the conditional probability that the state is $|\sqrt{n}\rangle$ when the measurement results $x_1$.


Now we evaluate the potentially leaked information to Eve in the sense of R\'enyi \cite{gpa}.
For the individual attacks, useful forms of R\'enyi information are known \cite{l1,A57,so}. 
According to \cite{A57}, for any binary pure states signal $\{ |\Psi_+ \rangle,|\Psi_-\rangle\}$, the maximum R\'enyi information gain is given by 
\begin{equation}
I_{opt}^R=\log_2 \left( 2-|\langle \Psi_+ |\Psi_-\rangle |^2 \right) \label{optr}.
\end{equation}

Since Eve can perform her measurement after she learns the parity of $m$, the problem is to find the maximum value of the R\'enyi information using a POVM on the binary signal $\{ | \sqrt{(1-\eta )n}  \rangle ,\ | -\sqrt{(1-\eta )n}  \rangle   \}$.
Thus substituting $|\Psi_\pm \rangle = |\pm \sqrt{(1-\eta )n} \rangle$ into Eq. (\ref{optr}), we obtain
\begin{equation}
I_{opt}^R( n,\eta )=\log_2\left( 2-\exp \left[ -\frac{ (1-\eta )n}{(\Delta x )^2} \right]\right) \label{ir-opt}.
\end{equation}

 Using expressions (\ref{s-gain}) and (\ref{ir-opt}) we obtain the secure key gain (with ideal error correction) \cite{l2, namiki1}:
\begin{eqnarray}
G(x_0,n,\eta)&=& \frac{1}{2 }\sum_{|x|>x_0}\textrm{Prob}(x)  \bigl( i_{AB} ( x,\eta n ) -I_{opt}^R (n,\eta ) \bigr) \nonumber  \\ &=& \sum_{x>x_0}\textrm{Prob}(x)  \bigl( i_{AB} ( x,\eta n ) -I_{opt}^R (n,\eta ) \bigr), \label{gainex}
\end{eqnarray}
where in the last expression we have used the properties, $\textrm{Prob}(x)=\textrm{Prob}(-x)$ and $i_{AB}(x,n)=i_{AB}(-x,n)$. Thus we will discuss taking $x \ge 0$ in what follows.

 Since, for given positive $n$, $0 \le I_{opt}^R <1$ and $i_{AB} $ is a monotonically increasing function of $x \ge 0 $ with $\lim _{x\to \infty} i_{AB} = 1 $, we can find $\tilde{x}$ which satisfies 
\begin{eqnarray}
i_{AB} ( \tilde{x},\eta n ) -I_{opt}^R (n,\eta) \ge 0.   \label{ineq}
\end{eqnarray}
From this inequality and expression (\ref{gainex}) with $\textrm{Prob}(x)\ge 0$, the choice of the threshold $x_0 \ge \tilde{x} $ gives $G > 0$. 

\begin{figure}[hbtp]
\epsfig{file=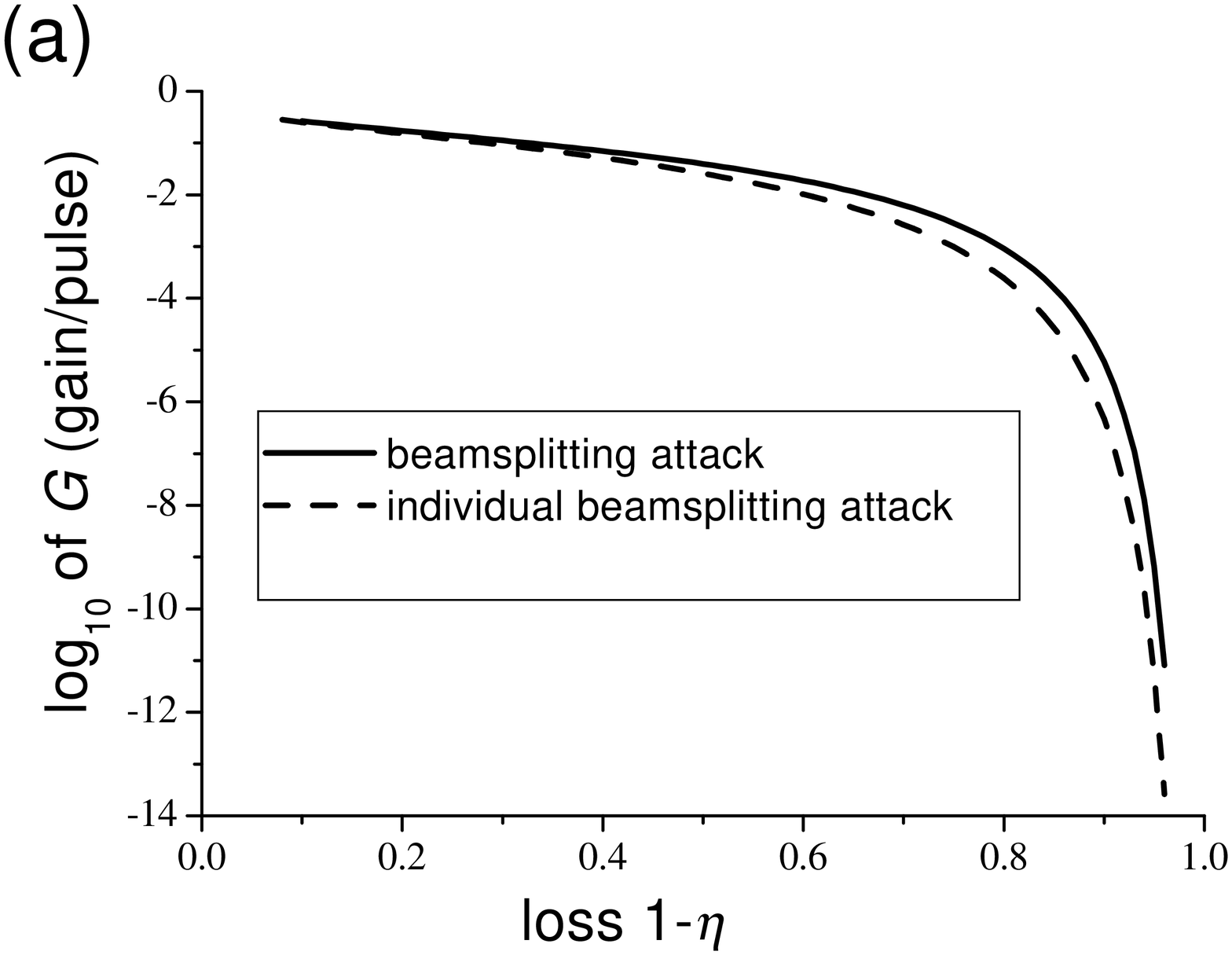,width=8.6cm}
\epsfig{file=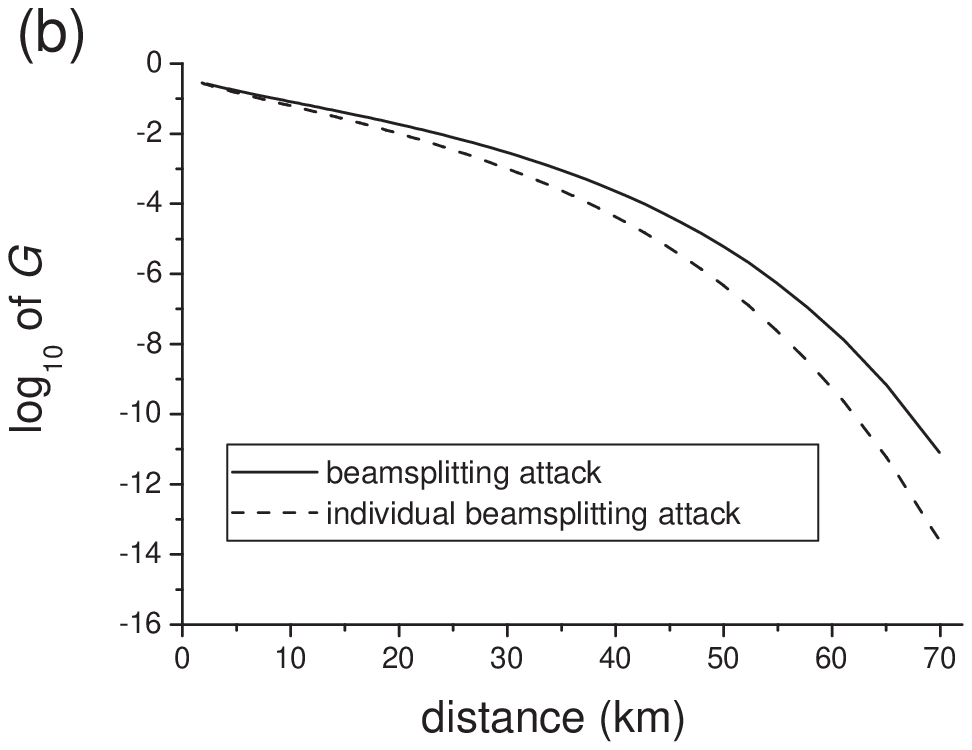,width=8.6cm}
\caption{ \label{test} The secure key gain $G$ is shown for given (a) loss $1- \eta $ and (b) distance in the absence of additional noise. The solid lines denote $G$ against a beam splitting attack where Eve performs quadrature measurement \cite{namiki1} and dashed lines denote $G$ against individual beam splitting attack. 
 Both of threshold $x_0$ and pulse intensity $n$ are optimized to maximize $G$. 
The distance is determined by using the optical fiber loss coefficient 0.2dB/km for 1.55$\mu$m wavelength. 
}
\end{figure}

 To obtain the maximum gain for a given $\eta$, $n$ and $x_0$ should be optimized simultaneously. A result of simultaneous optimization is shown in Fig. \ref{test}.

Thus far, we have considered only the amplitude damping of coherent states as experimental imperfections. Practically, noisy channel transforms an input coherent state into a mixture of coherent states. In the experiment \cite{hirano}, the observed quadrature distributions are Gaussian. Therefore we consider the channel which makes the Gaussian mixture:
\begin{equation}
| \alpha \rangle  \to  {\frac{\lambda}{\pi}} \int e^{- \lambda |\beta |^2}  | \alpha +\beta \rangle \langle \alpha +\beta|  d^2\beta , \ \lambda >0 \label{c-mixture}.
\end{equation}
In this case the observed phenomena for Bob is homogeneous broadening of the quadrature distributions.

If the observed quadrature distribution is Gaussian, Bob's information gain can be calculated by replacing $( \Delta x )^2$ with the observed quadrature variance $(\Delta x_{\textrm{obs}} )^2$ in Eqs. (\ref{pd}) and (\ref{replace}). However, the estimation of leaked information is not straightforward. In the beam splitting attack, $I^R$ is independent of $x$ and thus postselection is advantageous. In general, Eve's operation makes correlation between Bob's measurement result $x$ and the state Eve receives. In such case, $I^R$ depends on $x$ and thus the postselection is not necessarily advantageous. In the next paragraph we provide an extension of individual beam splitting attack which preserves the postselection advantage. 

\begin{figure}[tbhp]
\epsfig{file=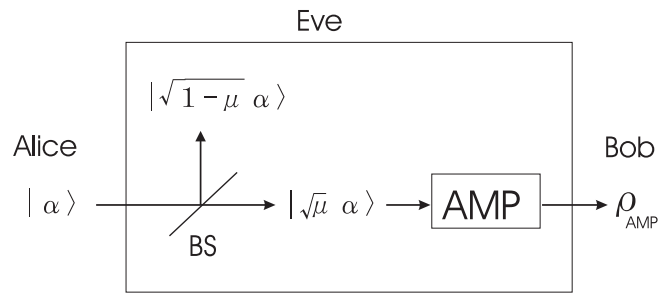,width=8.6cm}
\caption{ \label{amp-attack} Eve broadens Bob's observed quadrature variance $(\Delta x_{\textrm{obs}} )^2$ by combining a beam splitter (BS) and a phase-insensitive amplifier (AMP). } 
\end{figure} 

A lower bound of R\'enyi information for a given loss and Gaussian noise is given by considering the following attack: Eve performs a beam splitting with the reflectivity $1-\mu$ and then she operates a phase-insensitive amplifier \cite{amp} with the amplifier gain $g\ge 1$ (see Fig. \ref{amp-attack}). 
$\mu$ and $g $ are determined by the observed variance and mean photon number:
\begin{eqnarray}
(\Delta x_{\textrm{obs}} )^2=(2g-1)(\Delta x )^2 ,\\
g \mu n= \eta n .
\end{eqnarray}
This implies $\mu = \frac{2(\Delta x)^2}{(\Delta x_{\textrm{obs}} )^2+(\Delta x)^2} \eta =\frac{ \eta }{g} $ and Eve obtains more intense signal than that from individual beam splitting attack. By defining the effective loss $1- \mu$, the leaked information is estimated as the individual beam splitting attack: $I_{L}^R (n, \eta, g)=I_{opt}^R(n, \eta / g)$. 
In this case, we can still achieve $G> 0$ since $0 < I_{L}^R < 1 $ and $i_{AB} \to 1 \ (x \to \infty)$. Therefore transmission distance is unlimited by this attack.

\begin{figure}[htbp]
\epsfig{file=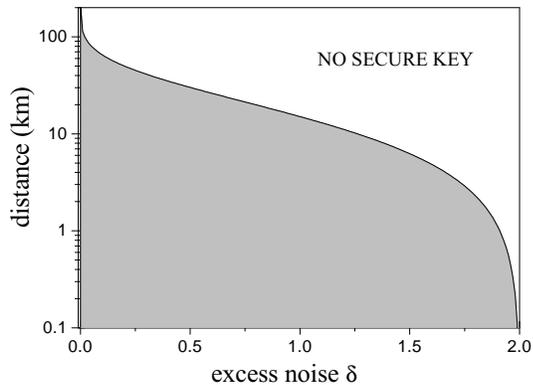,width=8.6cm}
\caption{A distance bound is shown as a function of normalized excess noise $\delta =\frac{(\Delta x_{\textrm{obs}} )^2}{(\Delta x )^2}-1 $. Out of the gray region $\delta \ge 2\eta$, secure key distribution is impossible. The distance is determined by using the optical fiber loss coefficient 0.2dB/km for 1.55$\mu$m wavelength. 
\label{d-bound}}
\end{figure} 

A distance bound in the presence of excess Gaussian noise is given by considering the following intercept-resend attack: Eve performs simultaneous measurement using a 50:50 beam splitter followed by two homodyne detectors and resends a coherent state whose amplitude is $\sqrt{2}$ times larger than the measured value of the simultaneous measurement. This operation is equal to continuous-variable quantum teleportation without EPR correlation \cite{c-tel}. So we refer it classical teleportation (CT).
The effect of CT is summarized as the expression (\ref{c-mixture}) with $\lambda = 1$: 
 \begin{equation}
 | \alpha \rangle  \to  {\frac{1}{\pi}} \int e^{-   |\beta |^2}  | \alpha +\beta \rangle \langle \alpha +\beta|  d^2\beta.
 \end {equation} 
The quadrature variance becomes 3 times larger than $(\Delta x )^2$ (2 is from simultaneous measurement and 1 is the variance of the resending coherent state). In each resending operation, Eve knows the state Bob receives and Bob has no information advantage, i.e., $I_{AB}<I_{E}$. 
 This condition is true for any coherent state scheme because CT is performed for every coherent state equally.  
Thus coherent state schemes are no longer secure if the observed excess noise is equal to $2(\Delta x )^2$ or larger.  

If we introduce loss after CT, the total signal transformation is
 \begin{equation}
 | \alpha \rangle  \to  {\frac{1}{\pi\eta  }} \int e^{-   \frac{|\beta |^2}{\eta }}  | \sqrt \eta \alpha +\beta \rangle \langle  \sqrt \eta \alpha +\beta|  d^2\beta.
 \end {equation} 
The observed variance is $(\Delta x_{\textrm{obs}} )^2=(1+2\eta )(\Delta x )^2$. The point is that the observed excess noise $2 \eta {(\Delta x )^2}$ becomes arbitrary small for high loss. Therefore there exist loss limit in the presence of any finite excess noise. By defining the normalized excess noise $\delta \equiv \frac{(\Delta x_{\textrm{obs}} )^2}{(\Delta x )^2}-1 $, a necessary condition of secure key distribution is given by $\delta < 2 \eta $ (see Fig.\ref{d-bound}).
This limitation is practical since realization of CT is possible within today's technology.

In conclusion, we have investigated the security of a continuous-variable quantum cryptographic scheme applying a postselection against individual beam splitting attack and an extension of this attack in the presence of excess Gaussian noise. It is shown that the secure key gain can be positive by setting a large postselection threshold and in this sense the transmission distance is unlimited as long as these attacks are concerned. 
We have also found a loss limit by a combination of a realistic intercept-resend attack and the loss mechanism which reduces excess noise on quadrature distributions. The limitation is given for every continuous-variable scheme using coherent states.

R.N. is supported by a scholarship from the Fujukai Foundation. 
This work was supported by ``R\&D on Quantum Commun. Tech."of MPHPT.
%



\end{document}